\documentclass[english,aps,showpacs,prc,superscriptaddress,preprintnumbers,floatfix,footinbib]{revtex4}
\usepackage{graphicx}

\begin{document}
\preprint{DAPNIA-06-51, JLAB-THY-06-477}
\title{Physical Origin of Density Dependent Force of the Skyrme Type 
within the Quark Meson Coupling Model}

\author{P.A.M. Guichon$^1$, H.H. Matevosyan$^{2,3}$, N. Sandulescu$^{1,4,5}$ and A.W. Thomas$^2$\\
$^1$SPhN-DAPNIA, CEA Saclay, F91191 Gif sur Yvette, France\\
$^2$Thomas Jefferson National Accelerator Facility,\\ 
12000 Jefferson Ave., Newport News, VA 23606, USA\\
$^3$Louisiana State University, Department of Physics \\
\& Astronomy, 202 Nicholson Hall, Tower Dr., LA 70803, USA\\
$^4$Institute of Physics and Nuclear Engineering, 76900 Bucharest, Romania\\
$^5$Service de Physique Nucleaire, CEA/DAM, F91680 Bruyeres le Chatel, France}

\begin{abstract}
A density dependent, effective nucleon-nucleon force of the 
Skyrme type is derived from the quark-meson coupling model 
-- a self-consistent, relativistic quark level description of 
nuclear matter. This new formulation requires no assumption 
that the mean scalar field is small and hence constitutes a 
significant advance over earlier work. The similarity of the 
effective interaction to the widely used SkM$^*$ force encourages
us to apply it to a wide range of nuclear problems, beginning 
with the binding energies and charge distributions of doubly 
magic nuclei. Finding acceptable results in this conventional 
arena, we apply the same effective interaction, within the 
Hartree-Fock-Bogoliubov approach, to the properties of 
nuclei far from stability. The resulting two neutron drip 
lines and shell quenching are quite satisfactory. Finally, 
we apply the relativistic formulation to the properties of 
dense nuclear matter in anticipation of future application
to the properties of neutron stars.
\end{abstract}
\maketitle

\section{Introduction}

Given that QCD is almost universally believed to be the fundamental 
theory of the strong interaction, one of the great challenges facing 
modern nuclear physics is to derive the properties of atomic nuclei from 
it. Amongst the hopes for a more fundamental description of nuclear matter
is to be able to extrapolate to extreme regions of density, 
asymmetry or even strangeness with rather more confidence than has 
hitherto been possible. One might also expect to understand the nuclear 
EMC effect~\cite{Geesaman:1995yd,Cloet:2005rt,Smith:2005ra} 
and a number of other challenges to traditional nuclear 
theory~\cite{Saito:2005rv,Strauch:2002wu}. 

While at the present time there has only been exploratory work aimed at 
relating QCD itself to nuclear 
structure~\cite{Thomas:2004iw,Chanfray:2005sa}, there has been 
considerably more work within the framework of quark models and effective 
field theories utilizing the symmetries of QCD. 
Here we 
shall concentrate on one particular model, the Quark-Meson Coupling (QMC) 
model~\cite{Guichon:1987jp,Guichon:1995ue,Saito:1995up,Saito:2005rv}, 
which is built around   
the self-consistent response of the relativistic, confined 
valence quarks to the Lorentz scalar mean field in nuclear matter. 
In previous work~\cite{Guichon:2004xg} we have shown that 
the main features of the effective
interactions used in nuclear physics can be quantitatively 
derived from the QMC model. In order to do this we rewrote 
the QMC model as an equivalent, effective Hamiltonian
with zero range many body interactions. The many body character of
the interactions was a direct consequence of the 
response of the quark structure to
the nuclear environment. We found that this Hamiltonian was
in quite good agreement with the popular 
Skyrme-III interaction~\cite{Vautherin:1971aw}, 
which led us to conclude that quark level dynamics does play a key role
in ordinary nuclear physics, even if it can be hidden within an effective 
many-body theory.

However, one unsatisfactory feature 
of the derivation in Ref.~\cite{Guichon:2004xg} 
was that the approximations which were required meant that 
one could not justify the use of the resulting  
effective Hamiltonian much beyond the density of ordinary nuclear 
matter --  $(\rho_{0}=0.16$fm$^{-3})$.
One of those approximations is actually tied to the physics of the model:
the response of the quark structure to the nuclear medium leads to
a non-linear meson field equation which is intractable in its full
generality. In Ref.~\cite{Guichon:2004xg} we relied on an iterative solution 
of the field equations, which essentially
amounted to an expansion in powers of the density. We showed that
this is an acceptable approximation for ordinary nuclear densities, with 
the medium effects then appearing as many body forces. Another approximation
used in that work was a non-relativistic expansion which, 
in the framework of the iterative
solution, appeared unavoidable. The latter approximation was also 
motivated by our wish 
to compare the effective Hamiltonian with the Skyrme interaction which is a
non-relativistic object. 
\footnote{This was also the motivation for the zero range approximation, which
could otherwise have been relaxed, 
at the price of algebraic complexity.}
These approximations, which were justified in the context of the  
earlier work, become untenable at higher densities. For instance, in
this model one finds that the velocity of sound exceeds the velocity of light
at about $3\rho_{0}.$ This is somewhat frustrating because it 
prevents us applying the model to the study of neutron stars. 

In this work we propose a new formulation which does not rely on the
approximations which we have just outlined. 
Instead of solving the meson field equation
iteratively, we expand the field around its nuclear expectation value
and treat the deviation as a perturbation. This leads to an effective
Hamiltonian with a density dependent two-body force. As the equation
for the expectation value can be solved exactly (at least in a numerical
sense), the bulk of the medium effects may be taken into account without
any assumption about the value of the density. This is a major advance 
with respect to the formulation in Ref.~\cite{Guichon:2004xg}. Moreover, 
the case of uniform
nuclear matter, which is the relevant approximation for neutron stars,
can now be treated in a way consistent with relativity. 

This paper is organized as follows.
In Section \ref{sec:The-QMC-model} we recall the basic elements of the QMC
model.
This is followed in Section \ref{sec:Hamiltonian} by the formal 
derivation of the equivalent, density dependent NN force.
A non-relativistic reduction of the full effective interaction 
is carried out in Section \ref{sec:Non-relativistic-expansion}  
in order to make a detailed comparison with the density dependent
Skyrme force in Section\ref{sec:Comparison-with-SkM}.
In Section \ref{sec:Hartree-Fock-calculations} we carry out a full Hartree-Fock calculation with the non-relativistic, effective force that we derived and 
compare the results for selected finite nuclei against experimental data.
Finally, in Section \ref{sec:High-density-uniform} we solve the full model
for uniform matter containing just nucleons -- leaving for future work 
the generalization of the method to include hyperons.
Section \ref{sec:Conclusion} presents some concluding remarks and 
suggestions for further work.

\section{\label{sec:The-QMC-model}The QMC model}

In the Quark Meson Coupling model, the nuclear system is represented
as a collection of confined clusters of three valence quarks. Because it 
is possible to carry through much of the calculation analytically, the 
MIT bag model was chosen to confine the quarks. 
It is assumed that the effect of having these bags overlap, that is 
clusters of more than three quarks, can be neglected.
This approximation certainly breaks
down at some density but we assume that it is still acceptable in the density
range that we consider. In this respect, we observe that the bag
model is an effective realisation of confinement, which must not be
taken too literally. Indeed, QCD 
lattice simulations~\cite{Alexandrou:2002sn,Bissey:2005sk}
strongly suggest that the true picture is closer to a Y-shaped color string
attached to the quarks. Outside this relatively thin string one has
the ordinary, non-perturbative QCD vacuum where the quarks from the other
nucleons can pass without disturbing the structure. Thus, while 
the bag model imposes a strict boundary condition which prevents the
quarks from travelling through its boundary, this must be seen as
the average representation of a more complex situation.  
One should not attribute a deep physical meaning to the boundary of
the cavity nor to its size. In particular, estimating the density at
which the non-overlap approximation breaks down as $\rho=1/V$, with
$V$ the bag volume, is certainly too pessimistic.

The salient feature of the QMC model is that the interactions are
generated by the exchange of mesons coupled locally to the quarks.
In a literal interpretation of the bag model, where only quarks and 
gluons can live inside the cavity, this coupling would be unnatural.
On the other hand, in terms of the more realistic picture presented by 
lattice QCD, the quarks
are attached to a string but otherwise move in the non-perturbative
QCD vacuum. There nothing prevents them from feeling the vacuum fluctuations
which we describe by meson fields. As in our earlier work, 
we limit our considerations
to the $\sigma,\,\omega$ and $\rho$ mesons. We recall that the $\sigma$
meson here is not the chiral partner of the pion. It is a chiral invariant
field which represents correlated two-pion exchange. The quarks of the
bag are in the Weinberg representation~\cite{Thomas:1981ps,Thomas:1982kv}, 
where the mass term $\bar{q}q$
is a chiral scalar which allows a chiral invariant coupling 
of the form $\sigma\bar{q}q$. 

Of course, this is not the end of the story
because the transformation to the Weinberg representation introduces
an infinite set of derivative pion-quark couplings. These produce
a complicated set of multi-pion exchanges, together with contact interactions.
We assume, in the spirit of both Quantum Hadrodynamics~\cite{Serot:1984ey} and 
the one boson exchange model, that the bulk of the corresponding physics
can be modeled by the $\sigma,\,\omega$ and $\rho$ exchanges between
quarks. It is clear that the long ranged single pion exchange does not 
fit in this picture and must be treated separately. However, because of 
its pseudoscalar nature, this exchange does not contribute to the
mean nuclear field. In the Hartree Fock approximation , which will
be our framework in the following, it contributes through its exchange
(Fock) term. This effect has been estimated in the framework of infinite
nuclear matter and found to be small. More exactly, most of the
effect can be absorbed by a readjustment of the couplings to the other
mesons. This is our first motivation for neglecting this part of the
interaction but we must keep in mind that the argument is only true
in the Hartree Fock approximation. When other nucleon correlations
are taken into account, the game changes but this is beyond the scope
of this work.

Our starting point is the \emph{classical} energy of a system of 
non-overlapping bags of quarks, which are coupled to the nuclear meson
fields $\sigma,\,\omega$ and $\rho$.
{}Following Ref.~\cite{Guichon:2004xg} we can write
\begin{equation}
E=\sum_{i} \left( \sqrt{P_{i}^{2}+M_{eff}(\sigma)^{2}}+
g_{\omega}\omega+V_{so} \right) +E_{mesons} \, ,\label{eq:1}\end{equation}
where the first term is the relativistic energy of nucleon $i$, with momentum
$P_i$ and effective mass $M_{eff}$, 
$g_{\omega}\omega$ is the repulsion felt by each nucleon as a 
result of its vector interaction with the time component of the 
$\omega$ mean field, 
$V_{so}$ is the spin-orbit potential and 
the static meson energy is 
\begin{equation}
E_{mesons}=\frac{1}{2}\int d\vec{r}\left[\left(\nabla\sigma\right)^{2}+m_{\sigma}^{2}\sigma^{2}\right]-\frac{1}{2}\int d\vec{r}\left[\left(\nabla\omega\right)^{2}+m_{\omega}^{2}\omega^{2}\right]\, ,\label{eq:2}\end{equation}
with $m_{\sigma},m_{\omega},m_{\rho}$ the meson masses. As usual,  
we consider only the time component of the vector fields. The effect
of the $\rho$, which can be treated by analogy with the $\omega$
field, will be introduced at the end. The expression for the spin orbit
interaction, $V_{so}$, first derived in Ref.~\cite{Guichon:1995ue},   
is not necessary at this stage.

The effective mass, $M_{eff}(\sigma)$, is the rest frame energy of
a quark bag in the $\sigma$ field of the medium, evaluated at the center of the
bag. In order to calculate $M_{eff}(\sigma)$ one needs 
to solve the bag equations for the relevant value of the 
$\sigma$ field. However, we have checked that
a quadratic expansion:
\begin{equation}
M_{eff}(\sigma)=M-g_{\sigma}\sigma+
\frac{d}{2}\left(g_{\sigma}\sigma\right)^{2}
\label{eq:3}
\end{equation}
is accurate up to values of $g_{\sigma}\sigma$ as large as 600MeV,
which corresponds to densities far beyond the physical limitations 
of the model. The parameter $d$ is the scalar polarizability of the
nucleon. It depends explicitly on the response of the quark structure
to the external scalar field. In the bag model it is well represented  
as a function of the bag radius, $R_B$, as 
\begin{equation}
d=0.0044+0.211R_{B}-0.0357R_{B}^{2}\, ,
\label{eq:4}
\end{equation}where both $d$ and $R_B$ are in fm.
This does not give exactly the values that were 
used in Ref.~\cite{Guichon:2004xg}, because
we have now included the contribution of the spin dependent,  
``hyperfine'' color interaction to the bag energy.
The coupling constant, $g_{\sigma}$, which by definition refers to
the nucleon, is related to the $\sigma$--quark coupling, 
$g_{\sigma}^{q}$, by
\begin{equation}
g_{\sigma}=3g_{\sigma}^{q}\int_{Bag}d\vec{r} \, \bar{q}q(\vec{r})\, ,
\label{eq:5}
\end{equation}
where $q$ is the valence quark wavefunction for the free bag. For the vector
couplings the relationship is 
\begin{equation}
g_{\omega}=3g_{\omega}^{q},\, g_{\rho}=g_{\rho}^{q}.
\label{eq:6}
\end{equation}
Equation (\ref{eq:3}) holds for any flavor, $f$, of hadrons, provided
the couplings and scalar polarizability are replaced by their corresponding
values $g_{\sigma}(f),d(f)$. In the bag model they are related to
$g_{\sigma}$ and $d$ in a well defined
way. Thus the treatment of flavors other than the nucleon is straightforward
and does not introduce new parameters. This matter will be developed
in forthcoming work. Here we limit our consideration to the case
where the medium contains just clusters with the quantum numbers 
of protons and neutrons.

\section{\label{sec:Hamiltonian}Hamiltonian}

By hypothesis, the meson fields are time independent. Therefore the
classical Hamiltonian for the nuclear system is simply
\begin{equation}
H(R_{i},P_{i})=E(R_{i},P_{i},\sigma\to\sigma_{sol},\omega\to\omega_{sol})\, ,
\label{eq:7}
\end{equation}
where $R_i$ and $P_i$ are the position and momentum of nucleon $i$ and  
$\sigma_{sol},\omega_{sol}$ are the the solutions of the equations
of motion corresponding to Eq.~(\ref{eq:1})
\[
\frac{\delta E}{\delta\sigma}=\frac{\delta E}{\delta\omega}=0\]
that is
\begin{eqnarray}
-\nabla^{2}\sigma+m_{\sigma}^{2}\sigma & = & -\sum_{i}\delta(\vec{r}-\vec{R}_{i})\frac{\partial}{\partial\sigma}\sqrt{P_{i}^{2}+M_{eff}(\sigma)^{2}},\label{eq:8}\\
-\nabla^{2}\omega+m_{\sigma}^{2}\omega & = & 
g_{\omega}\sum_{i}\delta(\vec{r}-\vec{R}_{i})\, .
\label{eq:9}
\end{eqnarray}
Note that in Eqs.~(\ref{eq:8},\ref{eq:9}) we have neglected the contribution
corresponding to the variation of the spin orbit interaction $V_{so}$.
As pointed out in Ref.~\cite{Guichon:2004xg}, 
this results in an error of order $V_{so}^{2}$
in the Hamiltonian and it is consistent to neglect it because the 
spin-orbit interaction in Eq.~(\ref{eq:1}) has been derived as a first
order perturbation~\cite{Guichon:1995ue}. 
Since the equation for the $\omega$ field
is linear, its solution is elementary and poses no new problem
with respect to our previous work. Its contribution to the energy
will be written later and we now concentrate on the field
equation for the $\sigma$.

We assume that it makes sense to write
\[
\sigma=<\sigma>+\delta\sigma\, ,\]
where the \emph{C-number} $<\sigma>$, also written $\bar{\sigma}$,
denotes the nuclear ground state expectation value, that is
\footnote{Of course this explicit expression is quite formal   
as it amounts to having
solved the problem.
}
\[
<\sigma(\vec{r})>=\int d\vec{R}_{1}...d\vec{R}_{A}\Phi^{*}(\vec{R}_{1}...\vec{R}_{A})\sigma(\vec{r},\vec{R}_{i},\vec{P}_{i})\Phi(\vec{R}_{1}...\vec{R}_{A})\]
and to consider the deviation, $\delta\sigma$, as a small quantity.
If we define
\begin{equation}
K=\sum_{i}\delta(\vec{r}-\vec{R}_{i})\sqrt{P_{i}^{2}+M_{eff}(\sigma)^{2}}\, ,
\label{eq:10}
\end{equation}
we see that the $\sigma$ field equation has the form
\begin{equation}
\left(-\nabla^{2}+m_{\sigma}^{2}\right)\left(\bar{\sigma}+\delta\sigma\right)=-\frac{\partial K}{\partial\sigma}=-\frac{\partial K}{\partial\sigma}(\bar{\sigma})-\delta\sigma\frac{\partial^{2}K}{\partial\sigma^{2}}(\bar{\sigma})-\cdots\label{eq:11}\end{equation}
We also expand $\frac{\partial K}{\partial\sigma}(\bar{\sigma}),\frac{\partial^{2}K}{\partial\sigma^{2}}(\bar{\sigma})$
about their expectation values
\begin{eqnarray}
\frac{\partial K}{\partial\sigma}(\bar{\sigma}) & = & <\frac{\partial K}{\partial\sigma}(\bar{\sigma})>+\delta\left[\frac{\partial K}{\partial\sigma}(\bar{\sigma})\right]\, ,\nonumber \\
\frac{\partial^{2}K}{\partial\sigma^{2}}(\bar{\sigma}) & = & <\frac{\partial^{2}K}{\partial\sigma^{2}}(\bar{\sigma})>+
\delta\left[\frac{\partial^{2}K}{\partial\sigma^{2}}(\bar{\sigma})\right]\, ,\label{eq:12}\end{eqnarray}
and suppose that 
\[
\delta\sigma,\,\delta\left[\frac{\partial K}{\partial\sigma}(\bar{\sigma})\right],\,\delta\left[\frac{\partial^{2}K}{\partial\sigma^{2}}(\bar{\sigma})\right]\]
are small quantities. We can then solve the meson field equation
order by order, which gives:
\begin{eqnarray}
\left(-\nabla^{2}+m_{\sigma}^{2}\right)\bar{\sigma} & = & 
-<\frac{\partial K}{\partial\sigma}(\bar{\sigma})>\label{eq:13}\\
\left(-\nabla^{2}+m_{\sigma}^{2}\right)\delta\sigma & = & 
-\delta\left[\frac{\partial K}{\partial\sigma}(\bar{\sigma})\right]-
\delta\sigma<\frac{\partial^{2}K}{\partial\sigma^{2}}(\bar{\sigma})>\, ,\nonumber \\
 & = & -\frac{\partial K}{\partial\sigma}(\bar{\sigma})+
<\frac{\partial K}{\partial\sigma}(\bar{\sigma})>-
\delta\sigma<\frac{\partial^{2}K}{\partial\sigma^{2}}(\bar{\sigma})>\, .
\label{eq:14}
\end{eqnarray}
As we limit the expansion of the Hamiltonian to order $(\delta\sigma)^{2},$
it is sufficient to solve the field equation at order $\delta\sigma$,
which corresponds to Eqs.~(\ref{eq:13},\ref{eq:14}). Using integration
by parts, the Hamiltonian (\ref{eq:1}) may be expanded as
\begin{eqnarray}
H & = & \int d\vec{r}\left.K\right|_{\bar{\sigma}}+\delta\sigma\frac{\partial K}{\partial\sigma}(\bar{\sigma})+\frac{1}{2}(\delta\sigma)^{2}\frac{\partial^{2}K}{\partial\sigma^{2}}(\bar{\sigma})\nonumber \\
 &  & +\frac{1}{2}\bar{\sigma}\left(-\nabla^{2}
+m_{\sigma}^{2}\right)\bar{\sigma}+\delta\sigma\left(-\nabla^{2}
+m_{\sigma}^{2}\right)\bar{\sigma}
+\frac{1}{2}\delta\sigma\left(-\nabla^{2}+m_{\sigma}^{2}\right)\delta\sigma\, .
\label{eq:15}
\end{eqnarray}
To this order we can replace
\[
\frac{\partial^{2}K}{\partial\sigma^{2}}(\bar{\sigma})\to<\frac{\partial^{2}K}{\partial\sigma^{2}}(\bar{\sigma})>\]
because this multiplies $(\delta\sigma)^{2}$. Using 
Eqs.~(\ref{eq:13},\ref{eq:14})
we then find:
\begin{equation}
H=\int d\vec{r}\,\left[K(\bar{\sigma})-\frac{1}{2}\bar{\sigma}<\frac{\partial K}{\partial\sigma}(\bar{\sigma})>+\frac{1}{2}\delta\sigma\left(\frac{\partial K}{\partial\sigma}(\bar{\sigma})-<\frac{\partial K}{\partial\sigma}(\bar{\sigma})>\right)\right].\label{eq:16}\end{equation}
Note that the mean field approximation amounts to neglecting $\delta\sigma$ in Eq.(\ref{eq:16}).
To complete the derivation we need a prescription for writing the quantum
form of $K$ and its derivatives. The important simplification is
that these are one body operators because they are evaluated at the
C-number point $\sigma=\bar{\sigma}$. Thus we can write
\[
K(\bar{\sigma})=\sum_{\alpha\beta}K_{\alpha\beta}(\bar{\sigma})a_{\alpha}^{\dagger}a_{\beta}\, ,\]
where $a_{\alpha}^{\dagger},a_{\alpha}$ are the creation and destruction
operators for the complete 1-body basis $|\alpha>$. The matrix elements
$K_{\alpha\beta}(\bar{\sigma})$ must be chosen so as to reproduce
the classical limit, Eq.(\ref{eq:10}). In the momentum space representation,
there is a natural choice  
\footnote{Spin and flavor labels are understood.}
\begin{equation}
K(\bar{\sigma})=\frac{1}{2V}\sum_{\vec{k},\vec{k}'}e^{i(\vec{k}
-\vec{k}').\vec{r}}\left(\sqrt{k^{2}+M_{eff}[\bar{\sigma}(\vec{r})]^{2}}
+\sqrt{k'^{2}+M_{eff}[\bar{\sigma}(\vec{r})]^{2}}\right)
a_{\vec{k}}^{\dagger}a_{\vec{k}'}\, ,
\label{eq:17}
\end{equation}
where the symmetrization is introduced to guaranty hemiticity and $V$
is the normalisation volume. We also choose 
\begin{equation}
\frac{\partial K}{\partial\sigma}(\bar{\sigma})=\frac{1}{2V}\sum_{\vec{k},\vec{k}'}e^{i(\vec{k}-\vec{k}').\vec{r}}\frac{\partial}{\partial\bar{\sigma}}\left(\sqrt{k^{2}+M_{eff}[\bar{\sigma}(\vec{r})]^{2}}+
\sqrt{k'^{2}+M_{eff}[\bar{\sigma}(\vec{r})]^{2}}\right)a_{\vec{k}}^{\dagger}a_{\vec{k}'}\, ,\label{eq:18}\end{equation}
with a similar expression for the second derivative. The ordering
ambiguities associated with products of non-commuting operators are fixed
by the normal ordering prescription, which amounts to removing that
part of the energy which originates from the interaction of one nucleon
with its own field.

%
%
The Hamiltoniam defined in Eq.~(\ref{eq:16}) is not a standard many-body 
problem, since we do not know $\bar{\sigma}$ and $\delta\sigma$
until the ground state nuclear wave function has been specified. 
Therefore the Hamiltonian must be determined (through $\bar{\sigma}$
and $\delta\sigma$ ) at each step of the self consistent procedure. 
This is a significant technical complication. On the other hand, for our  
purposes it is not necessary to
solve this Hamiltonian in the
general case. One of our main goals is to obtain the equation of state for
very dense nuclear matter, with a view to applying it to 
neutron stars~\cite{Lawley:2006ps},
but in this case we only need the uniform matter approximation for
which the model is easy to solve. Of course, we do need to consider finite size
effects for ordinary nuclei, a necessary step in order to show that
the model is realistic with respect to nuclear phenomenology. However,  
in this case, we can make a non-relativistic expansion and build a
density functional, $<H(\vec{r})>$, which can then be used in variational
calculations of nuclei.

\section{\label{sec:Non-relativistic-expansion}Non relativistic expansion}

We first consider the case of finite nuclei, in order to
verify that the model is indeed realistic. 
For this purpose we build the density functional, $<H(\vec{r})>$,  
using approximations which are standard in low energy nuclear physics.
The approximations that we use below involve a non-relativistic
expansion and the neglect of those velocity dependent and finite range forces
which involve more than 2 bodies -- as is the case for
conventional effective nuclear  
forces. This amounts to expanding the terms which involve either the
momentum or the gradient of the density in powers of the $\sigma$
nucleon coupling and stopping at order $g_{\sigma}^{2}$. To simplify
the expressions we omit the spin and flavor indices as long as they
are not truly necessary. We first define the number density, $D(\vec{r})$, 
and kinetic density, $\xi(\vec{r})$, by
\begin{equation}
D(\vec{r})=\frac{1}{V}\sum_{\vec{k},\vec{k}'}e^{i(\vec{k}-\vec{k}').\vec{r}}a_{\vec{k}}^{\dagger}a_{\vec{k}'},\,\,\,\,\xi(\vec{r})=\frac{1}{V}\sum_{\vec{k},\vec{k}'}e^{i(\vec{k}-\vec{k}').\vec{r}}\frac{k^{2}+k'^{2}}{2}a_{\vec{k}}^{\dagger}a_{\vec{k}'}.
\label{eq:20}
\end{equation}
Then, following the approximation scheme defined above and using
Eq.~(\ref{eq:3}) we find the following expression for the operator
$K$ and its derivatives
\begin{eqnarray}
\left.K\right|_{\bar{\sigma}} & = & D(\vec{r})M_{eff}[\bar{\sigma}(\vec{r})]+\frac{\xi(\vec{r})}{2M}\left(1+\frac{g_{\sigma}\bar{\sigma}}{M}\right)\, ,\nonumber \\
\left.\frac{\partial K}{\partial\sigma}\right|_{\bar{\sigma}} & = & D(\vec{r})\frac{\partial M_{eff}}{\partial\bar{\sigma}}+g_{\sigma}\frac{\xi(\vec{r})}{2M^{2}}\, ,\label{eq:21}\\
\left.\frac{\partial^{2}K}{\partial\sigma^{2}}\right|_{\bar{\sigma}} & = & dg_{\sigma}^{2}D(\vec{r})\, .\nonumber \end{eqnarray}
Substituting these expression into Eqs.~(\ref{eq:13},\ref{eq:14})  
and using the same approximations,  
we can solve for $\bar{\sigma}$ and $\delta\sigma$:
\begin{equation}
g_{\sigma}\bar{\sigma}=\frac{G_{\sigma}<D>}{1+
dG_{\sigma}<D>}-G_{\sigma}\frac{<\xi>}{2M^{2}}+
G_{\sigma}\frac{\nabla^{2}<D>}{m_{\sigma}^{2}}\, ,
\label{eq:22}
\end{equation}
\begin{equation}
\delta\sigma=\frac{1}{\tilde{m}_{\sigma}^{2}}\left(-\frac{\partial K}{\partial\sigma}(\bar{\sigma})+<\frac{\partial K}{\partial\sigma}(\bar{\sigma})>\right)+\frac{1}{\tilde{m}_{\sigma}^{2}}\nabla^{2}\frac{1}{\tilde{m}_{\sigma}^{2}}\left(-\frac{\partial K}{\partial\sigma}(\bar{\sigma})+<\frac{\partial K}{\partial\sigma}(\bar{\sigma})>\right)\, ,\label{eq:23}\end{equation}
where we have defined $G_{\sigma}=g_{\sigma}^{2}/m_{\sigma}^{2}$
and the (position dependent) effective $\sigma$ mass 
\begin{equation}
\tilde{m}_{\sigma}^{2}=m_{\sigma}^{2}\left(1+G_{\sigma}d<D>\right)\, .
\label{eq:24}
\end{equation}
We can now substitute Eqs.~(\ref{eq:21},\ref{eq:22},\ref{eq:23})
into Eq.~(\ref{eq:16}) to obtain the Hamiltonian of the model. In practice
we want the corresponding density functional in the Hartree Fock
approximation, so we need to evaluate (note that by definition 
$<\delta\sigma>=0)$
\begin{eqnarray}
<H(\vec{r})> & = & <K(\bar{\sigma})>-\frac{1}{2}\bar{\sigma}<\frac{\partial K}{\partial\sigma}(\bar{\sigma})>+\frac{1}{2}<\delta\sigma\frac{\partial K}{\partial\sigma}(\bar{\sigma})>\, ,\label{eq:25}\end{eqnarray}
where the ground state wave function is a Slater determinant, with
Fermi level $F$, built from the single particle 
wave functions $\Phi^{j}(\vec{r},\sigma,m)$.
We now restore the spin flavor dependence but restrict our considerations
to nuclei made of protons and neutrons. So the flavor index is just
the isospin projection $m=\pm1/2.$ As usual, we define~\cite{Vautherin:1971aw}~:
\begin{eqnarray}
\rho_{m}(\vec{r}) & = & \sum_{i\in F}\sum_{\sigma}
\left|\Phi^{i}(\vec{r},\sigma,m)\right|^{2},\,\,\,\,\rho(\vec{r})=
\sum_{m}\rho_{m}(\vec{r})\, ,
\label{eq:26}\\
\tau_{m}(\vec{r}) & = & \sum_{i\in F}\sum_{\sigma}
\left|\vec{\nabla}\Phi^{i*}(\vec{r},\sigma,mo)\right|^{2},\,\,\,\,\tau(\vec{r})=\sum_{m}\tau_{m}(\vec{r})\, ,
\label{eq:27}\\
\vec{J}_{m}(\vec{r}) & = & i\,\sum_{i\in F}
\sum_{\sigma\sigma'}\vec{\sigma}_{\sigma'\sigma}\times
\left[\vec{\nabla}\Phi^{i}(\vec{r},\sigma,m)\right]
\Phi^{i*}(\vec{r},\sigma',m),\,\,\,\,\vec{J}(\vec{r})=
\sum_{m}\vec{J}_{m}(\vec{r})\, ,
\label{eq:28}
\end{eqnarray}
and using standard techniques we find, after some algebra:
\begin{eqnarray}
<H(\vec{r})> & = & \rho M+\frac{\tau}{2M}\nonumber \\
 &  & +\frac{G_{\sigma}}{2M^{2}}\left(\rho\tau+\frac{1}{8}\sum_{m}\rho_{m}\nabla^{2}\rho_{m}\right)\nonumber \\
 &  & -\left(\frac{G_{\sigma}}{2m_{\sigma}^{2}}+\frac{G_{\sigma}}{4M^{2}}\right)\left(\rho\nabla^{2}\rho-\sum_{m}\left(\frac{1}{4}\rho_{m}\nabla^{2}\rho_{m}-\rho_{m}\tau_{m}\right)\right)\nonumber \\
 &  & -\frac{1}{2}\frac{G_{\sigma}}{1+dG_{\sigma}\rho}\left(\rho^{2}-\frac{1}{2(1+dG_{\sigma}\rho)^{2}}\sum_{m}\rho_{m}^{2}\right)\, .\label{eq:29}\end{eqnarray}
{}For completeness, we give the $\omega,\rho$ and spin orbit 
contributions (labelled $IS$ for isoscalar and $IV$ for isovector):
\begin{eqnarray}
<:{\cal H}_{so}^{IS}(\vec{r}):> & = & -\frac{1}{4M^{2}}\left[G_{\sigma}+\left(2\frac{\mu_{IS}}{\mu_{N}}-1\right)G_{\omega}\right]\sum_{mm'}\left[\left(1+\frac{1}{2}\delta_{mm'}\right)\rho_{m'}\vec{\nabla}.\vec{J}_{m'}\right]\, ,\label{eq:30}\\
<:{\cal H}_{so}^{IV}(\vec{r}):> & = & -\frac{G_{\rho}}{4M^{2}}\left[2\frac{\mu_{IV}}{\mu_{N}}-1\right]\sum_{mm'}\left[\left(mm'+\frac{1}{2}C_{mm'}\right)\rho_{m'}\vec{\nabla}.\vec{J}_{m}\right]\, ,\label{eq:31}\\
<:{\cal H}_{\omega}(\vec{r}):> & = & \frac{G_{\omega}}{2}\sum_{mm'}\left[\left(1-\frac{1}{2}\delta_{mm'}\right)\rho_{m}\rho_{m'}\right.\nonumber \\
 &  & \left.+\frac{1}{m_{\omega}^{2}}\left(\rho_{m}\nabla^{2}\rho_{m'}-\frac{1}{4}\delta_{mm'}\left(\rho_{m}\nabla^{2}\rho_{m}-4\rho_{m}\tau_{m}\right)\right)\right]\, ,\label{eq:32}\\
<:{\cal H}_{\rho}(\vec{r}):> & = & \frac{G_{\rho}}{2}\sum_{mm'}\left\{ \left(mm'-\frac{1}{2}C_{mm'}\right)\rho_{m}\rho_{m'}\right.\nonumber \\
 &  & \left.+\frac{1}{m_{\rho}^{2}}\left(mm'-\frac{1}{4}C_{mm'}\right)\rho_{m}\nabla^{2}\rho_{m'}+\frac{1}{m_{\rho}^{2}}C_{mm'}\rho_{m}\tau_{m'}\right\}\, , \label{eq:33}\end{eqnarray}
where: $C_{mm'}=\delta_{mm'}m^{2}+(\delta_{m,m'+1}+\delta_{m',m+1})/2$
and 
$G_{\omega}=g_{\omega}^{2}/m_{\omega}^{2},G_{\rho}=g_{\rho}^{2}/m_{\rho}^{2}.$
The isoscalar and isovector magnetic moments which appear in the spin
orbit interaction have the values
\begin{equation}
\mu_{IS}=\mu_{p}+\mu_{n}=0.88,\,\,\mu_{IV}=\mu_{p}-\mu_{n}=4.7\, .
\label{eq:34}
\end{equation}
As we have already pointed out, 
these expressions are the same as in our previous
work. We note that all terms which involve the
square of the spin density, $\vec{J}$, have been neglected. 
This is common practice and, 
since it amounts to treating the spin orbit interaction as a first
order perturbation, it is consistent with our derivation of the
expression for the effective mass, Eq.~(\ref{eq:1}).

It is convenient to have a more explicit expression for the density functional.
Using Eqs.~(\ref{eq:29} -- \ref{eq:33})
one can write, using the same notation as in Ref.~\cite{Chabanat:1997un}:
\begin{equation}
<H(\vec{r})>=\rho M+\frac{\tau}{2M}+{\cal H}_{0}+{\cal H}_{3}+
{\cal H}_{eff}+{\cal H}_{fin}+{\cal H}_{so}\, ,
\label{eq:35}
\end{equation}
where
\begin{eqnarray}
{\cal H}_{0}+{\cal H}_{3} & = & {\rho}^{2}\,\left[\frac{-3\,{G_{\rho}}}{32}+
\frac{{G_{\sigma}}}{8\,{\left(1+d\,\rho\,{G_{\sigma}}\right)}^{3}}-
\frac{{G_{\sigma}}}{2\,\left(1+d\,\rho\,{G_{\sigma}}\right)}+
\frac{3\,{G_{\omega}}}{8}\right]+\\
 &  & {\left({{\rho}_{n}}-{{\rho}_{p}}\right)}^{2}
\left[\frac{5\,{G_{\rho}}}{32}+\frac{{G_{\sigma}}}{8\,{\left(1+d\,\rho\,{G_{\sigma}}\right)}^{3}}-\frac{{G_{\omega}}}{8}\right],
\label{eq:40}\\
{\cal H}_{eff} & = & \left[\left(\frac{{G_{\rho}}}{8\,{m_{\rho}}^{2}}-\frac{{G_{\sigma}}}{2\,{m_{\sigma}}^{2}}+\frac{{G_{\omega}}}{2\,{m_{\omega}}^{2}}+
\frac{{G_{\sigma}}}{4\,{M_{N}}^{2}}\right)\,{{\rho}_{n}}+
\left(\frac{{G_{\rho}}}{4\,{m_{\rho}}^{2}}+
\frac{{G_{\sigma}}}{2\,{M_{N}}^{2}}\right)\,{{\rho}_{p}}\right]\,{{\tau}_{n}}
\nonumber \\
&  & +p\leftrightarrow n,
\label{eq:41}\\
{\cal H}_{fin} & = & \left[\left(\frac{3\,{G_{\rho}}}{32\,{m_{\rho}}^{2}}-\frac{3\,{G_{\sigma}}}{8\,{m_{\sigma}}^{2}}+\frac{3\,{G_{\omega}}}{8\,{m_{\omega}}^{2}}-\frac{{G_{\sigma}}}{8\,{M_{N}}^{2}}\right)\,{{\rho}_{n}}\right.\\
 &  & +\left.\left(\frac{-3\,{G_{\rho}}}{16\,{m_{\rho}}^{2}}-
\frac{{G_{\sigma}}}{2\,{m_{\sigma}}^{2}}+
\frac{{G_{\omega}}}{2\,{m_{\omega}}^{2}}-
\frac{{G_{\sigma}}}{4\,{M_{N}}^{2}}\right)\,{{\rho}_{p}}\right]{{\nabla}^{2}}({{\rho}_{n}})+p\leftrightarrow n,
\label{eq:42}\\
{\cal H}_{so} & = & \nabla\cdot{J_{n}}\left[\left(\frac{-3\,{G_{\sigma}}}{8\,{M_{N}}^{2}}-\frac{3\,{G_{\omega}}\,\left(-1+2\,{{\mu}_{s}}\right)}{8\,{M_{N}}^{2}}-\frac{3\,{G_{\rho}}\,\left(-1+2\,{{\mu}_{v}}\right)}{32\,{M_{N}}^{2}}\right)\,{{\rho}_{n}}\right.\\
 &  & +\left.\left(\frac{-{G_{\sigma}}}{4\,{M_{N}}^{2}}+\frac{{G_{\omega}}\,\left(1-2\,{{\mu}_{s}}\right)}{4\,{M_{N}}^{2}}\right)\,{{\rho}_{p}}\right]+p\leftrightarrow n.\label{eq:43}\end{eqnarray}

We have determined the couplings 
$G_{\sigma}, \, G_{\omega}, \, G_{\rho}$ by fixing the 
saturation density and binding energy of normal nuclear matter to be   
$\rho_{0}=0.16$fm$^{3}$ and  
$E_B=-15.85$ MeV, as well as the 
asymmetry energy of nuclear matter as $a_{4}=30$ MeV. Apart from 
a small readjustment
of the couplings, we found no significant sensitivity to the bag
radius. We therefore display our results for only one value, $R_{B}=0.8$fm,  
which is quite realistic~\cite{Thomas:1982kv}. 
The $\omega$ and $\rho$ masses are 
set at their experimental
values. The last parameter, which is not well fixed by experiment, is
the $\sigma$ mass. We shall use $m_{\sigma}=600,650,700,750MeV.$
The corresponding results are given in Table~\ref{cap:QMCparam}. 
We see that the incompressibility, 
$K_{N}$, is a little high with respect to the currently prefered range, 
but we point
out that this calculation has not yet taken into account the single pion
exchange interaction. We know~\cite{Guichon:2004xg} 
that the pion Fock term alone 
reduces $K_{N}$ by as much as 10\% and it is likely that this is
amplified by other correlations. 
\begin{table}
\begin{center}\begin{tabular}{|c|c|c|c|c|}
\hline 
$m_{\sigma}$(MeV)&
$G_{\sigma}$(fm$^{2})$&
$G_{\omega}$(fm$^{2})$&
$G_{\rho}$(fm$^{2})$&
$K_{N}$(MeV)\tabularnewline
\hline
\hline 
600&
12.652&
9.838&
9.67&
346\tabularnewline
\hline 
650&
12.428&
9.308&
8.583&
346\tabularnewline
\hline 
700&
12.254&
8.899&
7.724&
346\tabularnewline
\hline 
750&
12.116&
8.575&
7.048&
346\tabularnewline
\hline
\end{tabular}\end{center}

\caption{\label{cap:QMCparam}The couplings $G_{\sigma},G_{\omega},G_{\rho}$
for a bag radius $R_{B}=0.8$fm and several values of the scalar meson mass, 
$m_{\sigma}$. The last column
is the nuclear incompressibility.}
\end{table}

\section{\label{sec:Comparison-with-SkM}Comparison with the Skyrme force}

As a means of orientation, we compare the density functional of our model
with that corresponding to a typical density-dependent effective
interaction. We choose to compare with the popular Skyrme SkM$^*$ parametrisation rather than 
more recent ones, such  Sly4~\cite{Chabanat:1997un}, because in the latter case the formal
 identification is complicated  by  the larger  number of parameters. 
The SkM$^*$ interaction  depends on 6 parameters: $t_{0},t_{1},t_{2},t_{3},x_{0}$ and $W_{0}$
and its energy density~\cite{Chabanat:1997un} may be written, 
using the same notation
as in the previous section: 
\begin{eqnarray}
{\cal H}_{0}+{\cal H}_{3} & = & \frac{{\rho}^{\frac{1}{6}}\,{t_{3}}\,\left(2\,{\rho}^{2}-{{\rho}_{n}}^{2}-{{\rho}_{p}}^{2}\right)}{24}+\frac{{t_{0}}\,\left({\rho}^{2}\,\left(2+{x_{0}}\right)-\left(1+2\,{x_{0}}\right)\,\left({{\rho}_{n}}^{2}+{{\rho}_{p}}^{2}\right)\right)}{4}\, ,\label{eq:50}\\
{\cal H}_{eff} & = & \frac{\rho\,\tau\,\left(2\,{t_{1}}+2\,{t_{2}}\right)}{8}+\frac{\left(-{t_{1}}+{t_{2}}\right)\,\left({{\rho}_{n}}\,{{\tau}_{n}}+{{\rho}_{p}}\,{{\tau}_{p}}\right)}{8}\, ,\label{eq:51}\\
{\cal H}_{fin} & = & \frac{-\left(\rho\,\left(6\,{t_{1}}-2\,{t_{2}}\right)\,{{\nabla}^{2}}(\rho)\right)}{32}+\frac{\left(3\,{t_{1}}+{t_{2}}\right)\,\left({{\rho}_{n}}\,{{\nabla}^{2}}({{\rho}_{n}})+{{\rho}_{p}}\,{{\nabla}^{2}}({{\rho}_{p}})\right)}{32}\, ,\label{eq:52}\\
{\cal H}_{so} & = & \frac{-\left({W_{0}}\,\left(\rho\,\nabla\cdot J+
\nabla\cdot{J_{n}}\,{{\rho}_{n}}+
\nabla\cdot{J_{p}}\,{{\rho}_{p}}\right)\right)}{2}\, ,
\label{eq:53}
\end{eqnarray}
To simplify, we compare ${\cal H}_{eff},{\cal H}_{fin}$ and ${\cal H}_{so}$
with the QMC expressions Eqs.~(\ref{eq:41},\ref{eq:42},\ref{eq:43}), in
the case $N=Z.$ This detemines $t_{1},t_{2}$ and $W_{0}$. 
To find $t_{0},t_{3}$ and $x_{0}$,
we consider the term ${\cal H}_{0}+{\cal H}_{3}$. As the functional
form is not the same, we fit $\left({\cal H}_{0}+{\cal H}_{3}\right)_{QMC}$
with the form which appears in $\left({\cal H}_{0}+{\cal H}_{3}\right)_{SkM}$
in the range $\rho \in [0\to0.2$ fm$^{-3}]$. We first do this in the case
$N=Z$, which determines $t_{0}$ and $t_{3}$ and using these values we
do the fit for $Z/A=92/208$, which determines $x_{0}.$ The results
are collected in Table~\ref{cap:QMCvsSkM} for several values of the
$\sigma$ mass. We find satisfactory agreement with the SkM$^*$ parameters 
in the window $m_{\sigma}=650\div700$ MeV. This comparison suggests 
that our model may provide an acceptable representation 
of low energy nuclear physics.

However, this comparison is really rather qualitative. First, we have minimized
the isospin effects by setting $N=Z$ when determining the parameters
$t_{1},t_{2},W_{0}$. Second, the values we find for $t_{0}$ and $t_{3}$ depend
somewhat on the range of density we use for the fit. Thus a more direct
comparison of our model with actual nuclear data is desirable. This
is done in the next section.
\begin{table}
\begin{center}\begin{tabular}{|c||c|c|c|c|c|c||c|}
\hline 
$m_{\sigma}$(MeV)&
$t_{0}$(fm$^{2})$&
$t_{1}$(fm$^{4})$&
$t_{2}$(fm$^{4})$&
$t_{3}$(fm$^{5/2})$&
$x_{0}$&
$W_{0}$(fm$^{4})$&
Deviation \tabularnewline
\hline
\hline 
600&
-12.72&
2.64&
-1.12&
74.25&
0.17&
0.6&
33\%\tabularnewline
\hline 
650&
-12.48&
2.21&
-0.77&
71.73&
0.13&
0.56&
18\%\tabularnewline
\hline 
700&
-12.31&
1.88&
-0.49&
69.8&
0.1&
0.53&
18\%\tabularnewline
\hline 
750&
-12.18&
1.62&
-0.28&
68.28&
0.08&
0.51&
38\%\tabularnewline
\hline
\hline 
SkM$^*$&
-13.4&
2.08&
-0.68&
79&
0.09&
0.66&
0\%\tabularnewline
\hline
\end{tabular}\end{center}

\caption{\label{cap:QMCvsSkM}Comparison of the SkM$^*$ parameters with the QMC
predictions for several values of $m_{\sigma}.$}
\end{table}

\section{\label{sec:Hartree-Fock-calculations}Hartree Fock calculations for
finite nuclei}

As pointed out in the previous section, the QMC and the Skyrme energy
functionals have a similar structure. However, the density and the isospin
dependence of some terms, particularly $H_3$ and $H_{so}$, 
are rather different in
the two approaches. Thus, for the Skyrme functionals the  term $H_3$ has
a density dependence of the form $\rho^{\alpha}$. Originally this term
was interpreted as having been generated by a three-body contact interaction, 
which is
equivalent to a two-body density-dependent force 
in even-even nuclei~\cite{Vautherin:1971aw}.
In the QMC functional, the  density dependence of $H_3$ is much more 
complicated.
However, a more complicated density dependence could also appears in 
Skyrme type energy functionals if they are derived from a more fundamental theory 
based on G-matrix, such as density-matrix expansion (DME) method of Negele and Vautherin 
\cite{Negele72}. A comparison between the energy functionals corresponding
to DME and to standard Skyrme forces shows that in fact the strength of the 
density-dependent term is rather similar in both cases, in spite of their 
different density dependence.
It is interesting to observe that a density dependence of a 
fractional type, as appears in QMC, was previously considered in
phenomenological functionals by Fayans et al.~\cite{fayans}. The advantage of fractional expressions
in particle density is that the corresponding 
energy functional preserves causal behaviour up to very high densities.
This is actually the case for the QMC functional as well.

Another difference between the QMC and standard Skyrme energy 
functionals arises in the isospin dependence of the spin-orbit
term $H_{so}$. In both cases the form factor of the one-body spin-orbit
interaction for a nucleon with the isospin $m$ ($n$ or $p$) can
be written as:
\begin{equation}
 W_{m}(r)=a \nabla\rho_{m}+
b \nabla\rho_{\tilde{m}}\, ,
\end{equation}
where $\rho_m$ is the particle density, while $\tilde{m}$ denotes the
opposite isospin to $m$.
For the standard Skyrme forces the ratio $s=a/b$ is equal to 2. This isospin
dependence of the one-body spin-orbit potential is induced by the exchange term 
(since the Skyrme two-body spin-orbit force is isospin independent). 
For the QMC functional used in the calculations below, 
the ratio $s$ is about 2.78.
This strong isospin dependence in both the QMC and Skyrme functionals 
is in contrast
with the weak isospin dependence employed 
in the relativistic mean field models, 
in which the contribution of Fock (exchange) terms is neglected~\cite{ring}.

Starting from the QMC energy functional one can easily derive the
corresponding Hartree-Fock (HF) equations. They have a form similar to  
the Skyrme-HF equations, apart from the rearrangement term  and the one-body
spin-orbit interaction, which (as discussed above) have a different density and 
isospin dependence. 
The HF equations were solved in coordinate space, following 
the method described
in Ref.~\cite{Vautherin:1971aw} and the Coulomb interaction was treated 
in a standard way --  
i.e., the contribution of its exchange part was calculated 
in the Slater approximation.
The calculations were performed for the doubly magic nuclei 
$^{16}$O, $^{40}$Ca, 
$^{48}$Ca and $^{208}$Pb. For definiteness, the $\sigma$ meson mass 
has been set to $m_{\sigma}$=700MeV, as suggested by the comparison with 
the SkM$^*$ interaction.   At this point we recall that the QMC model 
is essentially classical because  both the position 
and velocity of the bag are assumed known in the energy 
expression  (\ref{eq:3}). The  quantization then leaves  some 
arbitrariness in the ordering of the momentum dependent pieces of the 
interaction. As pointed out in previous work~\cite{Guichon:2004xg}, in the 
non-relativistic approximation the difference between the orderings is 
equivalent to a change of about 10\% in  $m_{\sigma}$. 
In this work the ordering is fixed by the relativistic expression  
chosen for the operator $K$, Eq. (\ref{eq:17}). The non-relativistic 
reduction then  leads  to an ordering which is not the same 
as in Ref.~\cite{Guichon:2004xg}. This is  why the $\sigma$ meson mass 
that we use here is somewhat higher.  
Note that this ordering ambiguity is only of concern in 
the case of finite nuclei. 
In uniform matter, which is the relevant approximation for 
neutron stars, the problem does not exist.

The results for the binding energies and charge radii
are shown in Table~3.  
The charge densities are calculated with the proton form factor
usually employed in the Skyrme-HF calculations~\cite{Vautherin:1971aw}. 
{}From Table~3 one can
see that QMC-HF gives
results which are in reasonable agreement with the experimental values.
The agreement is not as good as that given by the recent Skyrme or RMF models,
but one should keep in mind that in these models the experimental
values for the binding energies and radii are included in the fitting procedure,
which is not the case for the QMC functional.

One also finds a similarly reasonable description for 
the spin-orbit splittings, shown in
Table~4. Since the isospin dependence in QMC-HF is stronger than in
Skyrme-HF, one would expect different values for nuclei
with large isospin asymetry. 
However, as can be seen from Table~4, the differences
are rather small. This is primarily because the spin-orbit splitting depends 
on the product of the spin-orbit form factor and the corresponding 
single-particle wave functions. Thus, if the wave functions are not strongly  
localised in the surface region, where $W_{\tau}(r)$ is effective, 
the influence
of the isospin dependence of $W_{\tau}(r)$ upon the splitting need 
not be so significant.

\begin{figure}[p]
\includegraphics[clip, scale=0.5]{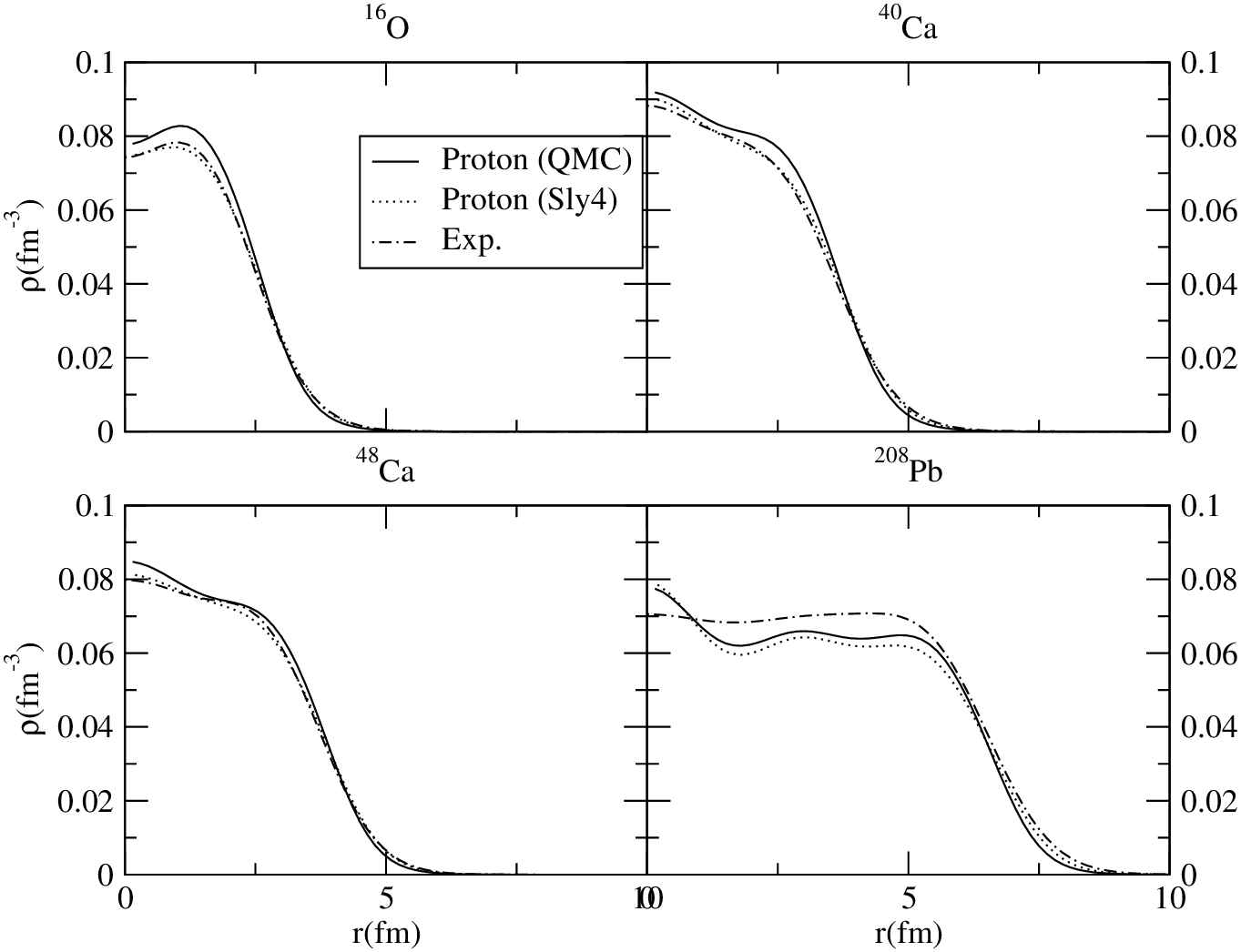}

\caption{\label{cap:PDensity}Proton densities of the QMC model compared with experiment and the prediction of the Skyrme Sly4 force.}
\end{figure}

\begin{figure}[p]
\includegraphics[clip, scale=0.5]{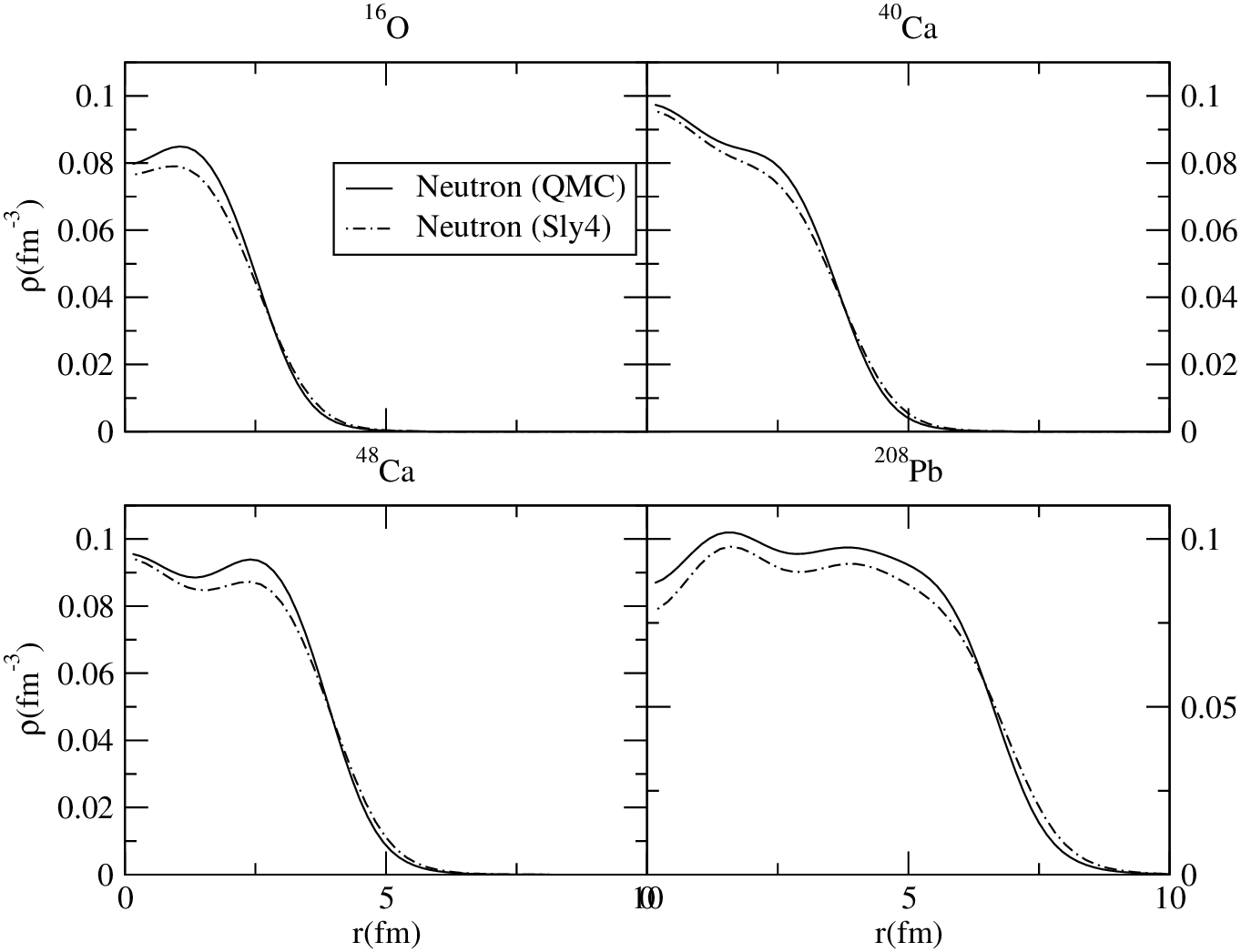}

\caption{\label{cap:NDensity}Neutron densities of the QMC model compared with the prediction of the Skyrme Sly4 force.}
\end{figure}

In Figs.~\ref{cap:PDensity},~\ref{cap:NDensity} we show the proton and neutron densities 
calculated with the QMC model 
and with the Sly4 Skyrme force~\cite{Chabanat:1997un}. In the proton case we also show the experimental values~\cite{DeJager:1987qc}. Once again the two models give 
rather similar results, with 
the largest
differences noticed for the neutron skin of $^{208}$Pb. More precisely,
{}for this nucleus the neutron skin 
(i.e., the difference between the neutron and proton
rms radii) is equal to 0.12fm for QMC, compared to 0.16fm for the
Skyrme force. With respect to experiment the QMC model 
tends to overestimate the density
in the central region, but the overall  agreement is quite good, given 
that the model has no parameter adjusted to fit the properties of finite nuclei.

Next we analyse the predictions of the QMC functional in nuclei far from 
stability. In order to do such a study, one must take into account 
the contribution of pairing correlations. 
They are treated here in the Hartree-Fock-Bogoliubov (HFB) approach.
For the pairing interaction we have taken a density-dependent contact 
interaction of the form~\cite{bertsch}:
\begin{equation}
V=V_0\left[1-\eta\left(\frac{\rho}{\rho_0}\right)^\alpha\right]\delta(r_1-r_2)\, .
\end{equation}
With this pairing force the QMC-HFB equations are local. 
They have been solved in
coordinate space by truncating the quasiparticle spectrum at an energy equal
to 60 MeV. The parameters of the pairing force have been fixed to the values:
$V_0=-333$ MeV fm$^{-3}$, $\eta=0.5$, $\alpha=1$, and $\rho_0=0.16$fm$^{-3}$,   
which were chosen to yield a reasonable average gap 
for the tin isotopes ( a benchmark for pairing models).

With this force we have investigated the position of 
the drip lines, which give the limits for bound nuclear systems.
Here we present the results for the two-neutron
drip line predicted by the QMC functional for Ni and Zr isotopes. 
The two-neutron
drip line is defined by the value of N (number of neutrons) for 
which the two-neutron
separation energy becomes negative. 
In the QMC-HFB calculations the neutron drip line appears 
around N=60 for Ni and around N=82 for Zr. 
These positions of the drip line are similar to 
the predictions provided by the Skyrme force SLy4, 
commonly used in non-relativistic
calculations~\cite{Chabanat:1997un}.

Another important property which characterizes nuclei far from 
stability is the 
shell quenching, which has important consequences for 
the astrophysical rapid capture 
processes. The phenomenological energy functionals show that 
when one approaches
the drip line the shell gaps associated with the magic numbers of 
neutrons or protons are 
quenched in comparison with stable nuclei. 
The shell quenching can be tested by calculating the
two-neutron separation energies, $S_{2n}$, across a magic number. 
Here we present the QMC-HFB results
for the neutron magic number N=28. 
The variation of $S_{2n}$ across N=28 is calculated
for two extreme values of proton numbers, 
namely Z=32 (proton drip-line region) and Z=14 
(neutron drip-line region). One thus finds that $S_{2n}$ changes by about 8 MeV for
Z=32 and by about 2-3 MeV for Z=14. 
This strong shell quenching is very close to that  
obtained in the Skyrme-HFB calculations 
(see Fig.~15 of Ref.~\cite{Chabanat:1997un} ).

In conclusion, the QMC functional gives a reasonable description of 
the known double magic
nuclei and predicts, for nuclei far from stability,   
similar properties to those found in mean field
Skyrme models.
\begin{table}[tp]
\begin{center}
\begin{tabular}{|c||c|c||c|c|}
\hline 
&
$E_B$ (MeV, exp)&
$E_B$ (MeV, QMC)&
$r_{c}$ (fm, exp)&
$r_{c}$ (fm, QMC)\tabularnewline
\hline
\hline 
$^{16}O$&
7.976&
7.618&
2.73&
2.702\tabularnewline
\hline 
$^{40}Ca$&
8.551&
8.213&
3.485&
3.415\tabularnewline
\hline 
$^{48}Ca$&
8.666&
8.343&
3.484&
3.468\tabularnewline
\hline 
$^{208}Pb$&
7.867&
7.515&
5.5&
5.42\tabularnewline
\hline
\end{tabular}
\end{center}

\caption{\label{cap:EsurAandR}Binding energy and radii calculated in QMC-HF, 
as described in the text.}
\end{table}
\begin{table}[t]
\begin{center}
\begin{tabular}{|c||c|c||c|c|}
\hline 
&
Neutrons (exp)&
Neutrons (QMC)&
Protons (exp)&
Protons (QMC)\tabularnewline
\hline
\hline 
$^{16}O,\,1p_{1/2}-1p_{3/2}$&
6.10&
6.01&
6.3&
5.9\tabularnewline
\hline 
$^{40}Ca,\,1d_{3/2}-1d_{5/2}$&
6.15&
6.41&
6.00&
6.24\tabularnewline
\hline 
$^{48}Ca,\,1d_{3/2}-1d_{5/2}$&
6.05 (Sly4)&
5.64&
6.06 (Sly4)&
5.59\tabularnewline
\hline 
$^{208}Pb,\,2d_{3/2}-2d_{5/2}$&
2.15 (Sly4)&
2.04&
1.87 (Sly4)&
1.74\tabularnewline

\hline
\end{tabular}
\end{center}

\caption{\label{cap:Splittings}Comparison between the QMC and ``experimental'' 
spin orbit splittings. 
Because the experimental splittings are no so well known in 
the case of $^{48}Ca$ and $^{208}Pb$, we give the values  
corresponding to the Skyrme Sly4 prediction. }
\end{table}

\section{\label{sec:High-density-uniform}High density uniform matter}
In the previous section we have shown that the non-relativistic approximation
to the Hamiltonian developed in section \ref{sec:Hamiltonian} gives
an acceptable phenomenology for ordinary nuclei. This give us some
confidence to examine the consequences of the model in the high density
region, relevant to the neutron star problem. The classical application of 
a density dependent force of the Skyrme type to this problem was that 
of Negele and Vautherin~\cite{Negele:1971vb}. Of course, all non-relativistic 
effective theories must break down as the density increases and we therefore 
concentrate on a truly relativistic treatment in this case. 
{}For neutron stars one can
restrict consideration 
to uniform nuclear matter, which implies that $<K(\bar{\sigma})>$
and its derivatives are numbers independent of $\vec{r.}$ The constant
expectation value of the sigma field $\bar{\sigma}$ is then determined
by the self consistent equation (see Eq.~(\ref{eq:13})):
\begin{equation}
\bar{\sigma}=-\frac{1}{m_{\sigma}^{2}}<\frac{\partial K}{\partial\sigma}(\bar{\sigma})>\, ,
\label{eq:100}
\end{equation}
which is solved numerically. The fluctuation, $\delta\bar{\sigma}$,
is given by Eq.~(\ref{eq:14}), where the effective 
$\sigma$ mass, $\tilde{m}_{\sigma}$,  
is now constant and the solution, expressed in terms of the $\sigma$-meson 
Green function, ${\cal G}_\sigma$, is 
\begin{eqnarray}
\delta\sigma(\vec{r}) & = & \int d\vec{r}'{\cal G}_{\sigma}(\vec{r}-\vec{r}')\left(-\frac{\partial K}{\partial\sigma}(\bar{\sigma},\vec{r}')+<\frac{\partial K}{\partial\sigma}(\bar{\sigma})>\right)\nonumber \\
 & = & \int d\vec{r}'\frac{d\vec{q}}{(2\pi)^{3}}\frac{e^{i\vec{q}.(\vec{r}-\vec{r}')}}{q^{2}+\tilde{m}_{\sigma}^{2}}\left(-\frac{\partial K}{\partial\sigma}(\bar{\sigma},\vec{r}')+<\frac{\partial K}{\partial\sigma}(\bar{\sigma})>\right)\label{eq:103}\end{eqnarray}
Note that, even though $\bar{\sigma}$ is constant, the operator $K$
and its derivative have an explicit dependence on position (see
Eqs.~(\ref{eq:17},\ref{eq:18})) which disappears only in the expectation
value. 

For uniform matter the energy density is 
\begin{equation}
\epsilon=\frac{<H>}{V}=\frac{1}{V}\int d\vec{r}\,\left[<K(\bar{\sigma})>
-\frac{1}{2}\bar{\sigma}<\frac{\partial K}{\partial\sigma}(\bar{\sigma})>
+\frac{1}{2}<\delta\sigma(\vec{r})\,
\frac{\partial K}{\partial\sigma}(\bar{\sigma},\vec{r})>\right] \, .
\label{eq:104}
\end{equation}
In a Fermi gas we have
\begin{equation}
<K_{m}(\bar{\sigma})>=\frac{2}{(2\pi)^3)}\int_{o}^{k_{F}(m)}d\vec{k}\sqrt{k^{2}+M^{2}(\bar{\sigma})},\,\,\,\,
<K(\bar{\sigma})>=\sum_{m}<K_{m}(\bar{\sigma})>\, ,\label{eq:105}
\end{equation}
with $k_{F}(m)$ the Fermi level of the isospin species $m.$ Using
the solution for $\delta\sigma$ we get, using standard techniques
:
\begin{eqnarray}
\frac{<H>}{V} & = & <K(\bar{\sigma})>+\frac{1}{2m_{\sigma}^{2}}\left(<\frac{\partial K}{\partial\sigma}(\bar{\sigma})>\right)^{2}\nonumber \\
 & + & \frac{1}{(2\pi)^{6}}\sum_{f}\int_{0}^{k_{F}(m)}d\vec{k}_{1}d\vec{k}_{2}\frac{1}{(\vec{k}_{1}-\vec{k}_{2})^{2}+\tilde{m}_{\sigma}^{2}}\frac{\partial}{\partial\bar{\sigma}}\sqrt{k_{1}^{2}+M^{2}(\bar{\sigma})}\frac{\partial}{\partial\bar{\sigma}}\sqrt{k_{2}^{2}+M^{2}(\bar{\sigma})}.\label{eq:106}\end{eqnarray}
{}Finally we must add the contributions $<V_{\omega}>,<V_{\rho}>$ associated 
with $\omega$ and $\rho$ exchange. However, since these interactions are
purely 2-body there is nothing new with respect to our previous calculation.
One finds 
\begin{eqnarray}
\frac{<V_{\omega}>}{V} & = & \frac{G_{\omega}}{2}\rho^{2}-G_{\omega}\sum_{m}\frac{1}{(2\pi)^{6}}\int_{0}^{k_{F}(m)}d\vec{k}_{1}d\vec{k}_{2}\frac{m_{\omega}^{2}}{(\vec{k}_{1}-\vec{k}_{2})^{2}+m_{\omega}^{2}},\label{eq:107}\\
\frac{<V_{\rho}>}{V} & = & \frac{G_{\rho}}{8}\left(\rho_{p}-\rho_{n}\right)^{2}-G_{\rho}\sum_{mm'}C_{mm'}\frac{1}{(2\pi)^{6}}\int_{0}^{k_{F}(m)}d\vec{k}_{1}\int_{0}^{k_{F}(m')}d\vec{k}_{2}\frac{m_{\rho}^{2}}{(\vec{k}_{1}-\vec{k}_{2})^{2}+m_{\rho}^{2}}.\label{eq:108}\end{eqnarray}

The energy density (\ref{eq:106},\ref{eq:107},\ref{eq:108}) and
its non-relativistic version, given in 
Section \ref{sec:Non-relativistic-expansion}, 
correspond to the same model. However, as in any theory, the parameters
depend on the approximation scheme and therefore we need to readjust
the couplings $G_{\sigma},G_{\omega},G_{\rho}$ to reproduce the saturation
properties for normal nuclear matter. 
We do this for $R_{B}=0.8$fm and $m_{\sigma}=700MeV$.
As can be seen in Table~\ref{cap:Rcouplings}, the couplings we get
are significantly different from those obtained with the non-relativistic
expansion. The most important effect arises from the non-relativistic
approximation (\ref{eq:22}) to the self consistentency equation (\ref{eq:100})
for $\bar{\sigma}$. If we were to use the same coupling constants
in both equations, we would find a value of $\bar{\sigma}$ which,
at the saturation point, differs by 15\%. 

\begin{table}
\begin{center}\begin{tabular}{|c||c|c|c|c|}
\hline 
&
$G_{\sigma}$(fm$^{2})$&
$G_{\omega}$(fm$^{2})$&
$G_{\rho}$(fm$^{2})$&
$K_{N}$(MeV)\tabularnewline
\hline
\hline 
Non relativistic&
12.254&
8.899&
7.724&
346\tabularnewline
\hline 
Relativistic&
11.334&
7.274&
4.56&
342\tabularnewline
\hline
\end{tabular}\end{center}

\caption{\label{cap:Rcouplings}The couplings $G_{\sigma},G_{\omega},G_{\rho}$
of the non-relativistic (first line) and the relativistic (second
line) version of the model for $R_{B}=0.8$fm and the $m_{\sigma}=700MeV$}
\end{table}

\begin{figure}
\begin{center}\includegraphics[clip, scale=0.5]{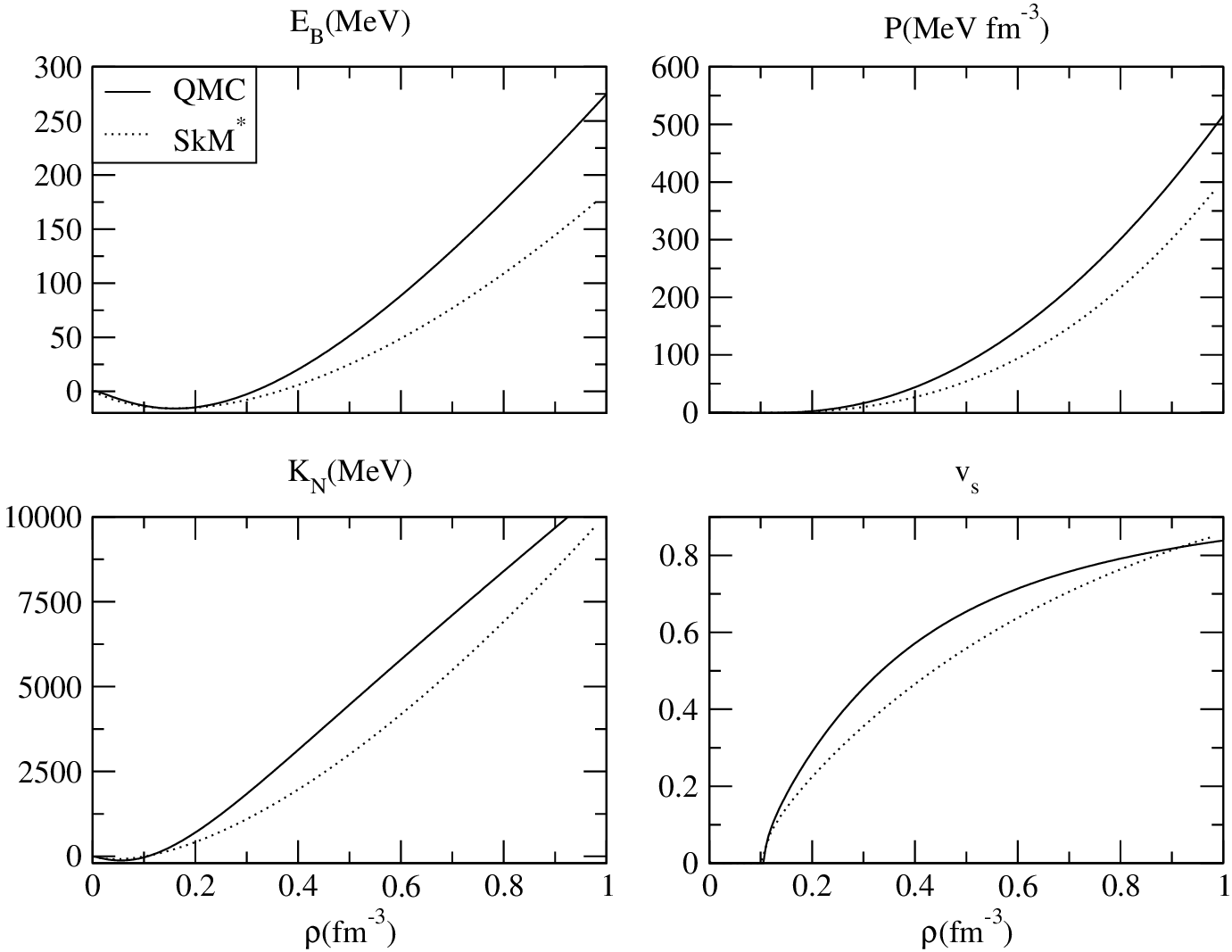}\end{center}
\caption{\label{cap:Sym-Matter}Comparison of the QMC and SkM$^*$ model for symmetric
matter.}

\end{figure}

In Fig.~\ref{cap:Sym-Matter} we show the binding energy $E_B$, pressure $P$, incompressibility $K_{N}$ and sound velocity $V_{s}$ of symmetric nuclear matter
predicted by our model (QMC) and the Skyrme (SkM$^*$) interaction.  We recall the definitions:
\begin{eqnarray}
E_B=\frac {\epsilon}{\rho}-M,\ P=\rho^2\frac{\partial E_B}{\partial\rho},
\ K_N=9\frac{\partial P}{\partial\rho},\ V_s=\sqrt{\frac{\rho K}{9(\epsilon+P)}}\, .
\end{eqnarray}
As we have restricted the flavor composition to protons and neutrons,
the curves must not be taken too seriously for densities above $2\div3$
times $\rho_{0}$, because there the appearance of hyperons 
is expected to become important.
However, the comparison between QMC and SkM$^*$ is still meaningful.
One sees that the behavior of the two models is qualitatively
similar in the range of densities displayed in the plots. Of course,  
at much higher density the SkM$^*$ model would violate the causality limit
$(V_s=1)$, because of its non-relativistic nature. By contrast the
QMC model, because of its relativistic formulation, gives $V_s<1$
at any density. The QMC binding energy curve is stiffer than that for 
SkM$^*$. This is not unexpected in view of the high value of $K_{N}(\rho_{0})$
given by QMC. However, the difference is not spectacular when one considers
the full density range. This is even more obvious when one looks at
the incompressibility curve as a function of density. In fact the
curves cross somewhere above $\rho=1$fm$^{-3}$. 

\begin{figure}
\begin{center}\includegraphics[clip, scale=0.5]{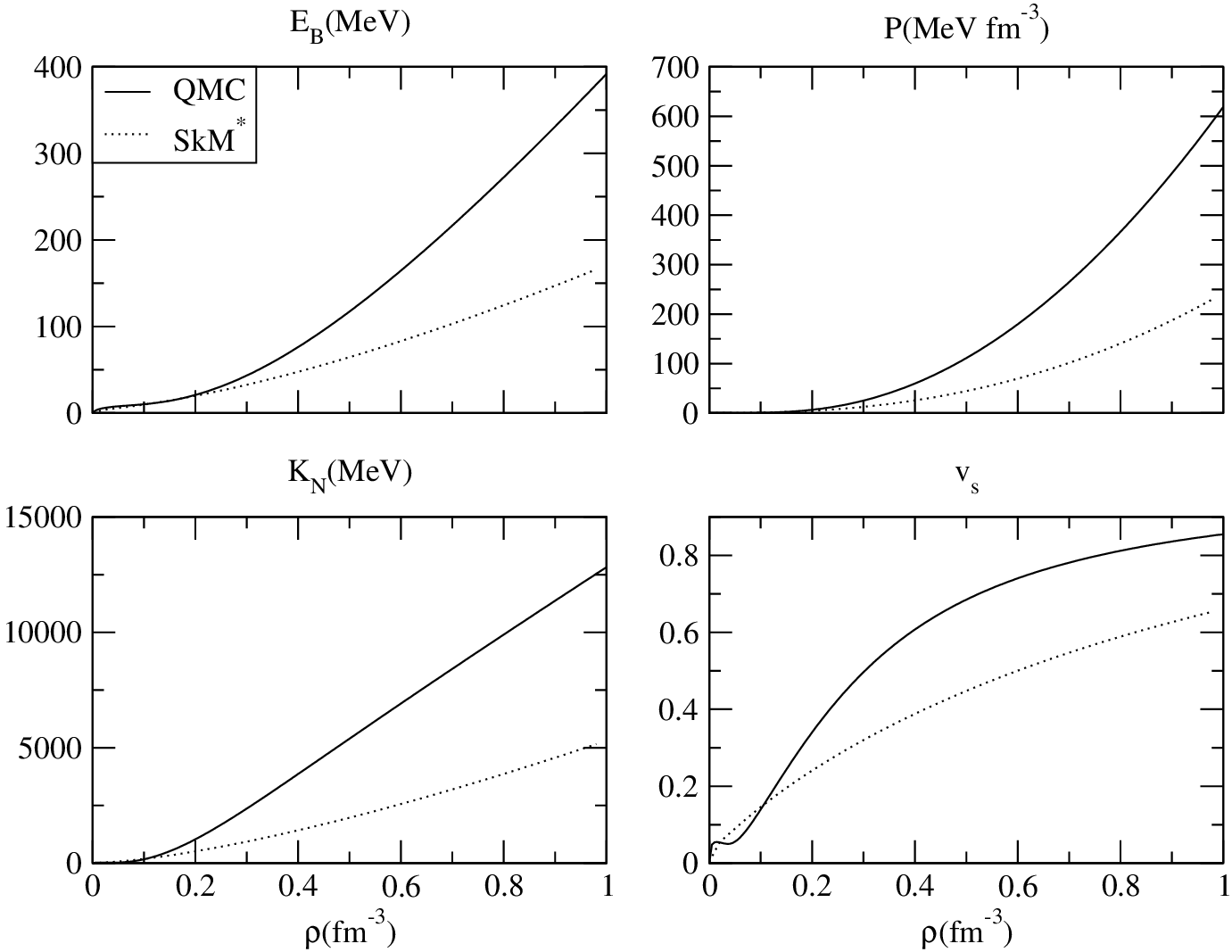}
\end{center}

\caption{\label{cap:NeutronMatter}Comparison of the QMC and SkM$^*$ model for
neutron matter.}

\end{figure}

The situation changes radically when one considers the case of neutron
matter -- as shown in Fig.~\ref{cap:NeutronMatter}. The QMC equation
of state is considerably stiffer than that for SkM$^*$ above $\rho\sim0.2$fm$^{-3}.$
This difference is not too surprising 
because in the Skyrme interaction
the isospin dependence appears only in the 2-body interaction through
the $x_{0}$ parameter, while in the QMC model it is also generated
by the non-linearity of the $\sigma$ field equation, as can be seen
in Eq.(\ref{eq:40})
\footnote{Note that this equation can only be used at moderate density since
it is a non-relativistic approximation. %
}. Thus one may expect that the neutron stars predicted by the QMC model
will have higher maximum masses than in the SkM$^*$ model. However one
cannot draw firm conclusions until the effect of hyperons has been
calculated.

\section{\label{sec:Conclusion}Conclusion}
Starting from the QMC model, we have derived a density 
dependent effective nucleon-nucleon force which is not 
limited to low density. The application of this effective 
interaction to the properties of doubly magic nuclei and then
to nuclei far from stability yields results in quite acceptable 
agreement with popular effective forces which are far more 
phenomenological and which have rather more parameters. 
This provides one with 
confidence in the underlying physical concepts of 
the QMC model, most notably the scalar polarizability of 
a hadron which results from the self-consistent response of 
the confined valence quarks to the mean scalar field in the 
medium. An important corollary is that one can immediately 
apply the model to derive appropriate effective hyperon-nucleon
effective forces as a function of density with no additional 
parameters. Given the lack of experimental guidance on 
anything other than the $\Lambda -N$ force at relatively low 
density and the apparent importance of hyperons in dense matter,
this will be our next application of the methods developed here. 
We remain mindful that our present work has been built upon the MIT 
bag model, which involves a relatively crude characterization of 
confinement. To the extent that the effect of internal structure is 
controlled by the scalar polarizability this is unimportant. Nevertheless, it 
will also be interesting to see the effect of using more sophisticated 
quark models.

To put our results in perspective, 
we have applied the present approach to a relativistic Hartree-Fock
treatment of dense matter consisting of nucleons. In the case of symmetric matter we find an equation
 of state which is not very different from the one predicted by the Skm$^*$ force but in the case of neutron matter
it is definitively stiffer. However  to infer the implications  for  the 
neutron stars properties we need to 
investigate the changes induced when hyperons are allowed.  Finally, 
given that we are starting with a quark model for the nucleon 
itself, it will be extremely interesting find the point at which 
a phase transition to quark matter becomes favourable. In this work we have formulated  the QMC model 
so that it can be used formally at any density but it is clear that it must break at some point. 
Whether the central density in the massive neutron stars is high enough to reach this point is a tantalizing question. 

\section*{Acknowledgements}
This work was supported by the {\it Espace de Structure Nucl\'eaire 
Th\'eorique du CEA}, by  NSF grant  NSF-PHY-0500291 and by DOE  grant DE-AC05-84150, under which SURA operates  Jefferson Laboratory.  We thank J. Rikovska-Stone for stimulating discussions.

\end{document}